\documentclass[10pt,journal,compsoc]{IEEEtran}
\newif\ifpeerreview

\peerreviewfalse

\usepackage[nocompress]{cite}
\usepackage{url}
\usepackage{amsmath,amssymb,graphicx}

\usepackage{lipsum} % Only used to generate random text.

\usepackage{xcolor}%imkl

\usepackage[switch]{lineno}

\newcommand{\paperID}{XXXX}

\title{Spectra2pix: Generating Nanostructure \\ Images from Spectra}

\usepackage{authblk}
\author[1,3]{Itzik Malkiel}
\author[2]{Michael Mrejen}
\author[1,4]{Lior Wolf}
\author[2]{Haim Suchowski}
\affil[1]{Department of Computer Science, Tel Aviv University}
\affil[2]{School of Physics and Astronomy, Tel Aviv University}
\affil[3]{Microsoft}
\affil[4]{Facebook AI Research}

\begin{document}

\IEEEtitleabstractindextext{%
\begin{abstract}
%{\color{red}The past decade has witnessed the advent of nano-photonics where the light-matter interaction is shaped, almost at-will, with man-made, designed nanostructures TOO COMPLEX AND UNCLEAR; CONTRADICTS THE NEXT SENTENCE WITH THE ALMSOT AL WILL}. 
The design of the nanostructures that are used in the field of nano-photonics has remained complex, very often relying on the intuition and expertise of the designer, ultimately limiting the reach and penetration of this groundbreaking approach. Recently, there has been an increasing number of studies suggesting to apply Machine Learning techniques for the design of nanostructures. Most of these studies engage Deep Learning techniques, which entails training a Deep Neural Network (DNN) to approximate the highly non-linear function of the underlying physical process between spectra and nanostructures. At the end of the training, the DNN allows an on-demand design of nanostructures, i.e. the model can infer nanostructure geometries for desired spectra. In this work, we introduce spectra2pix, which is a model DNN trained to generate 2D images of the designed nanostructures. Our model architecture is not limited to a closed set of nanostructure shapes, and can be trained for the design of any geometry. We show, for the first time, a successful generalization ability by designing a completely unseen sub-family of geometries. This generalization capability highlights the importance of our model architecture, and allows higher applicability for real-world design problems.
\end{abstract}

}

% Make Title
\ifpeerreview
\linenumbers \linenumbersep 15pt\relax 
\author{Paper ID \paperID\IEEEcompsocitemizethanks{\IEEEcompsocthanksitem This paper is under review for ICCP 2020 and the PAMI special issue on computational photography. Do not distribute.}}
\markboth{Anonymous ICCP 2020 submission ID \paperID}%
{}
\fi
\maketitle

% The first section title should be wrapped inside a \IEEEraisesectionheading as follows.
\IEEEraisesectionheading{
  \section{Introduction}\label{sec:introduction}
}
% The very first letter of the paper is a 2 line initial drop letter
% followed by the rest of the first word in caps.
% 
% form to use if the first word consists of a single letter:
% \IEEEPARstart{A}{demo} file is ....
% 
% form to use if you need the single drop letter followed by
% normal text (unknown if ever used by the IEEE):
% \IEEEPARstart{A}{}demo file is ....
% 
% Some journals put the first two words in caps:
% \IEEEPARstart{T}{his demo} file is ....
% 
% Here we have the typical use of a "T" for an initial drop letter
% and "HIS" in caps to complete the first word.

\begin{figure*}[h]
\includegraphics[width=1.0\linewidth]{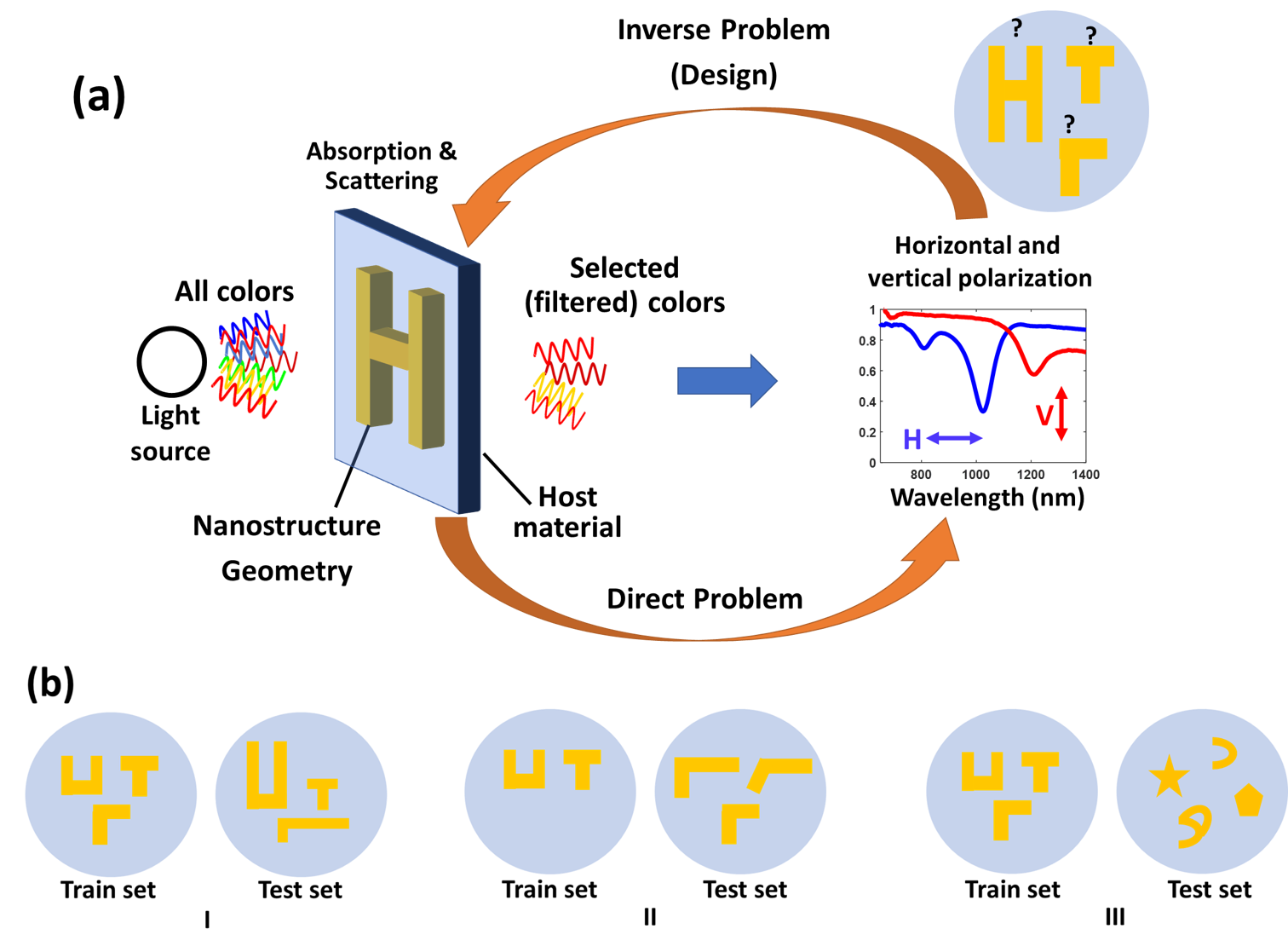}
\centering
\caption{Deep learning Nano-photonics. (a) Interaction of light with plasmonic nanostructures. Incoming electromagnetic radiation interacts with nanostructure in a resonant manner, leading to an effective optical response. The spectra of both polarizations are dictated by the geometry of the nanostructure, rather than the chemical composition.
(b) the different categories of generalization as explained in the main text.}
\label{fig:intro}
\end{figure*}

\IEEEPARstart{I}{nverse} design of nanophotonics structures, i.e. obtaining a geometry for a desired photonic function, has been a challenge for decades. Due to the highly nonlinear nature of this optimization problem, it requires, , when applying evolutionary or topology optimization algorithms, hundreds to several thousands of iterations for a single design task. Recently, modern machine learning algorithms have been applied to the inverse problem in nanophotonics and demonstrated great promise.

%{\color{blue}polish the next 3 paragraphs}
% for both polarizations of the electromagnetic field,
The interaction of light with nano-scale material, embedded in dielectric, can be characterized by various properties of the outgoing light \cite{somRef4}. Figure \ref{fig:intro}a illustrates such optical response, for which a white light (containing all the colors) interacts with a metallic subwavelength geometry. This interaction results with partial transmission due to absorption and scattering. The partial transmission entails that these nano-scale geometries cannot be observed by a conventional microscope. This property is also known as the diffraction limit, which stipulates that optical information smaller than roughly half the illumination wavelength is not retrievable. 

Predicting the optical response of a nanostructure geometry requires solving the full set of Maxwell equations. This problem, denoted by 'direct problem' in figure \ref{fig:intro}a, is also considered as the more feasible problem, and can be solved via simulations. The more challenging direction is the 'inverse problem' of inferring the nanoscale geometry from a measured or desired spectra. 

The major contributions that have been published so far to design nanostructures by utilizing machine learning techniques, can be categorized into three categories as far as the designed structures are concerned. The first, and the most fundamental one, is obtaining a model that is capable to design nanostructures from the same shape and material it was trained on, but with different properties, such as sizes, angles, host material and so on. Work such as \cite{peurifoy2018nanophotonic, sajedian2019finding, liu2018training, ma2018deep, malkiel2018plasmonic, malkiel2018deep} fall within this category where the general structure is maintained (eight alternating shells particle or m alternating layer of thin films) and the ML algorithm works to provide optimized parameters of the structure. The second category incorporates models that are able to generalize and designing geometries with shapes that differ from the set of shapes used during training, but are still considered to be in the same family,  i.e. the model can generalize to other shapes that are similar but not identical to the set of shapes it was trained on. For example, in this work, we showcase that our model can infer a “L” shaped nanostructure, given matched spectra, while the model was trained on different shapes, such as “H”, “h”, “n”, etc. Additional attempts to devise such a model have been recently presented in \cite{liu2018generative}, where the authros tried to test the generalization ability of a model by training a model on a set of digist, leaving one digit as a test set, however in their work, the model designed a shape from the set of shapes it was trained on, and didn't seem to generalize as expected. The third category incorporates models that are able to design any geometry, with any shape, achieving the ultimate generalization capability. The generalization ability of such models should be verified via a proper holdout test set, comprising structures sampled from a completely different distribution the model was trained on. To this end, studies that argue to provide a model that is able to design nanostructures for any spectra, should put extra care in constructing a test set that would verify the generalization level of the model at hand.

The above three categories, illustrated in Figure \ref{fig:intro}b, are ordered by the complexity of the underlying physical problem. Whereas the most desirable capability is of course the latest category, which can design any geometry with any shape. 

To verify such a property, one may need to harvest a synthetic dataset with great diversity that spans the entire distribution of supported geometries. However such a dataest is not yet available in the community, and may take hundreds of simulation hours to harvest. In this work, we utilize the dataset from \cite{malkiel2018deep, malkiel2018plasmonic}, and use it to veirfy that our suggested model can generalize at least to the level of category two described above.

Under the context of the model at hand and the assumption that a large volume of data is available for learning, the first immediate step to achieve such a generic capability is to design a model, that have enough degrees of freedom to allow the design of any geometry.

Previous work \cite{peurifoy2018nanophotonic, sajedian2019finding, liu2018training, malkiel2018deep, malkiel2018plasmonic, malkiel2017deep}, introduced a model that can be classified under the first category above, i.e. the model is able to infer geometries of the same or similar shapes it was trained on, which have variable sizes, angels and epsilon host materials. However, in order to be able to design any geometry, one should allow a larger degree of freedom. Specifically, in \cite{malkiel2018deep, malkiel2018plasmonic} the model architecture was designed to retrive coding vectors that encode the geometry shape to the H shape family. To somehow circumvent the inherent limitation of this encoding, further degrees of freedom are obtained as the authors asked the model to predict each edge presence, the length of the edge and the angle between the inner edge and the top right edge. 

As we look to expand these capabilities to the second or even third more desirable categories, perhaps the most direct way to allow a model to design any geometry, is to adopt an architecture that supports generating any shape. 

Recent work in computer vision has suggested pix2pix\cite{isola2017image}, a neural based model that learns to map images from source domain to target domain. Given an input image, the model learns to generate images according to some ground truth image labels. Applied on different types of datasets, pix2pix showcases the ability of neural networks to generate realistic images, that preserve different types of underlying logic, such as mapping gray images to color images, facades labels to images, maps to aerial, and more. 

In this work, we adapt a few key properties from the pix2pix architecture. Our model receives spectra as an input, and retrieves a 2D image, for which the pixels forms the designed nanostructure geometry shape. Hence, we name our method spectra2pix. 

Spectra2pix model aims to expand the capabilities of previous work to the second or even third more desirable categories. The model focuses on solving the inverse problem of inferring a nanostructure geometry from a given spectra and material properties. Differently from the previous bi-directional model \cite{malkiel2018deep, malkiel2018plasmonic}, the spectra2pix architecture supports the generation of any geometry, by training the model to regress the raw pixel values of the 2D images of the geometries at hand. The training task is being enforced by optimizing the spectra2pix model to minimize a pixelwise loss term, applied on the generated image with the ground truth image. In this work, we have published a new version of the dataset introduce in \cite{malkiel2017deep, malkiel2018deep, malkiel2018plasmonic}, incorporating the 2D images of the geometries. The dataset and the code can be found on \url{https://github.com/ItzikMalkiel/spectra2pix}. 
%{\color{blue} make it anonymous for ICCP}

\begin{figure*}[t]
\includegraphics[width=1.0\linewidth]{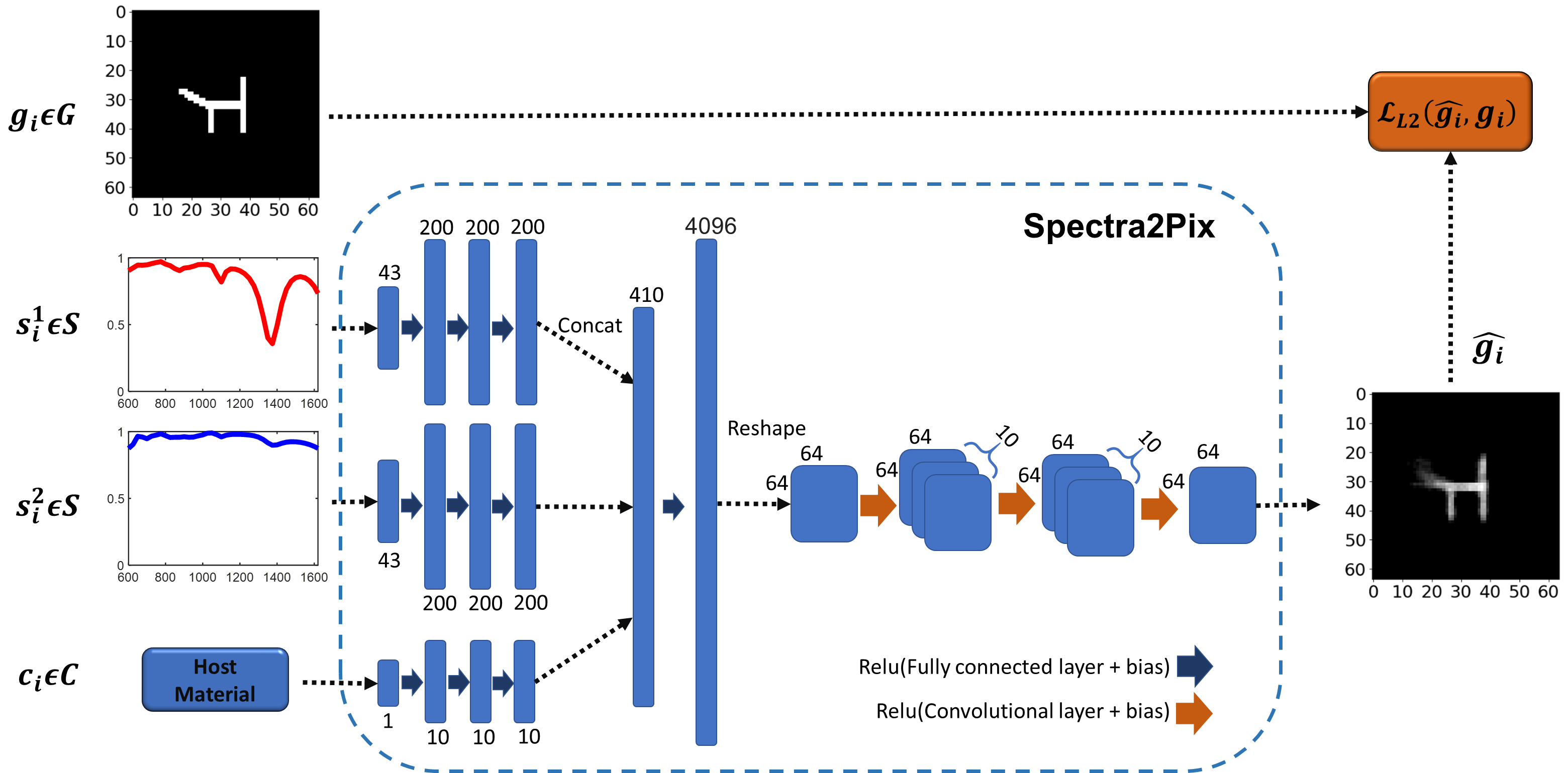}
\centering
\caption{ Spectra2Pix architecture. The model receives two spectra and epsilon host value. Each data forward through three parallel fully connected layers (with separate weights), then concatenated into one intermediate vector. Next, another fully connected is applied that transform the vector into a higher dimension (equals to 64X64). The latter is reshaped to a matrix, which then applied through 3 convolutional layers, resulting with a 64X64 matrix.}
\label{fig:architecture}
\end{figure*}

Our contribution is three-fold: (1) We introduce spectra2pix, a model that conceptually can design any 2D nanostructure geometry. (2) We are the first to report a successful generalization ability of the model, exhibiting the design of geometries sampled from a fairly different distribution that the model was trained on. Which is associated with level 2 described above. (3) We transform the dataset from \cite{malkiel2017deep, malkiel2018deep, malkiel2018plasmonic} to 2D image representation, and publish the new version to the community.

In the method section, we present the model architecture and training setting. Next, in the results section, we show qualitative results of our spectra2pix model, showcasing the generalization ability of our model. The results we show here hold great potential for the more general goal which will, however, require more extensive, broader, and more generic dataset. This can be addressed in further research where, possibly, the learning dataset could be crowdsourced from the ML nanophotonics growing community.

\section{Related Work}

\begin{figure*}[t]
\includegraphics[width=1.0\linewidth]{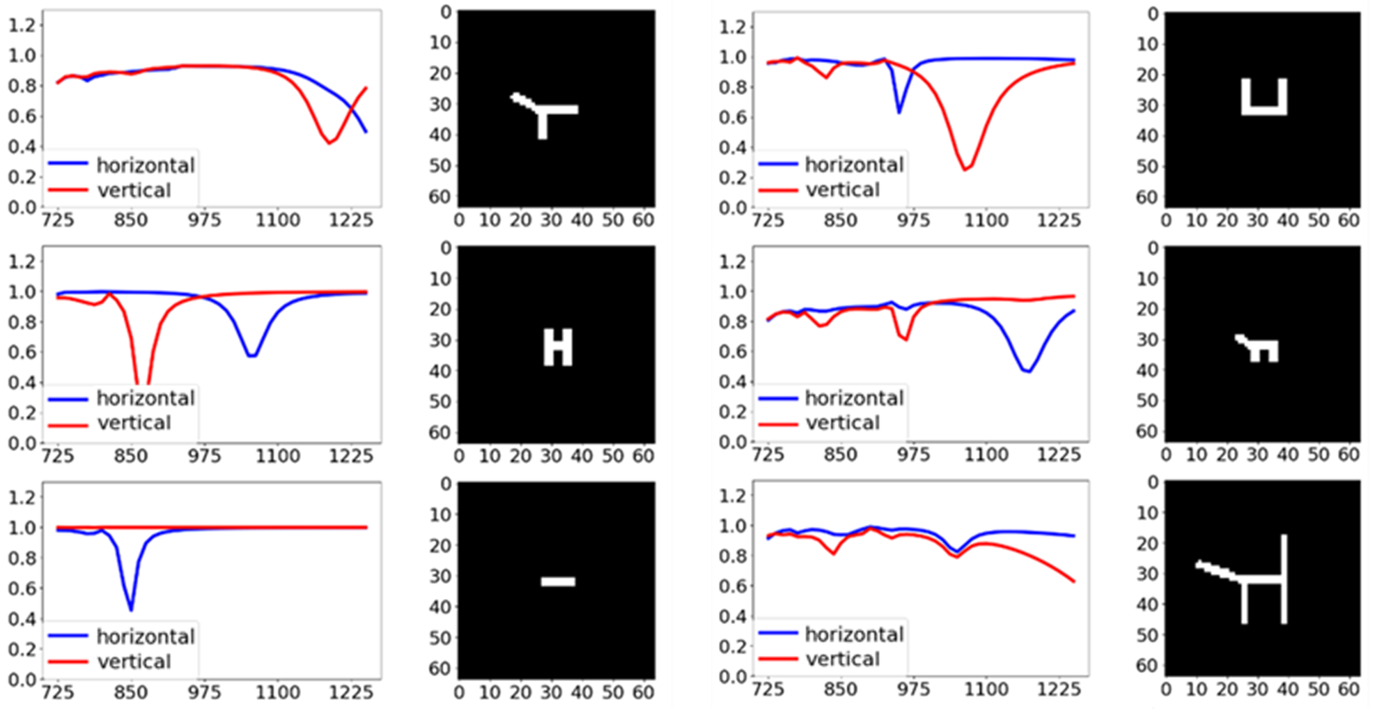}
\centering
\caption{ Six samples from the train dataset. We see the wide variety of spectra and geometries span by the encoding of the “H” edges dimensions and presence. The spectra correspond to horizontal and vertical polarizations in transmission. In our train set, all experiments made of gold, each with a different host dielectric, varying in the range [1.0,3.0] (not shown in the figure).}
\label{fig:dataset}
\end{figure*}

Malkiel at el \cite{malkiel2017deep, malkiel2018deep, malkiel2018plasmonic} introduced the first neural based model for the design of nanostructures. In this work, the authors proposed a bi-directional neural network architecture, which is able to solve both the inverse problem of designing nanostructure and the direct problem of inferring the optical characteristics of the designed geometry. The advantages of the bi-directional model are twofold. First, a bi-directional model is able to streamline the design process, by retrieving an immediate prediction for the optical properties of the designed nanostructure. That way, the designer can match the desired spectra with the recovered spectra, which can be used also in understanding the confidence level of the model for the specific design. Second, a bi-directional model allows co-adaptation between the networks of both directions, leading to better robustness and higher stability for the predictions. The introduced model was trained on synthetic data centered around different variants of the H shape, and was also applied on measured spectra form nano-fabricated materials conducted in lab. This model architecture is inherently limited to H family. This is to date the only fabricated experimental demonstration of the geometry prediction capability of deep learning network.

In the same studies above \cite{malkiel2017deep, malkiel2018deep, malkiel2018plasmonic}, the authors showcase the ability of the model to infer geometries of the same or similar shapes it was trained on, which have variable sizes, angles and epsilon host materials. This experiment corresponds to category one, as described above. However, in order to design any geometry of any shape, one should allow a larger degree of freedom. Specifically, the bi-directional architecture was designed to retrieve coding vectors that encode the geometry shape of the “H” family. %To somehow circumvent the inherent limitation of this encoding, further degrees of freedom were obtained as the  asked the model to predict each edge presence, the length of the edge and the angle between the inner edge and the top right edge.

In \cite{liu2018generative} the authors propose a generative adversarial network (GAN) for generating 2D nanostructure images from spectra. The authors created a synthetic dataset of geometries associated with multiple families, such as squares, circles, sectors, crosses and shapes from the MNIST dataset (which incorporate handwritten digits). Then, the authors showcased the ability of the model to randomly design test samples from each one of the families above, using a model that was trained with samples from all families. This evaluation corresponds to category one presented above, as the model task is to infer geometries from the same template it already saw in the training (this time, only with different attributes such as sizes, angles etc). 

During the second evaluation described in  \cite{liu2018generative}, the authors tested a higher level of generalization, which correlates to the second category described above. In this evaluation, the authors used a holdout test set comprises of a complete sub-family set of geometries. Specifically, the authors decided to keep all the samples that corresponds to digit “5” from the MNIST family in a holdout test set, and trained their model on the rest of the dataset. As reported in \cite{liu2018generative}, the topologies of the predicted geometry and the ground truth geometry differ considerably (the predicted geometry composed a variation of the digit ‘3’), but the overall spectra of the predicted geometry possess somehow similar features to the input spectra, with some discrepancies in few specific locations. In addition, the authors also argue that without GAN training, their model collapses, and generates images of random pixels. When optimizing an inverse function of a single network, one can often obtain a solution that satisfies the inversion criteria, but which, however, does not create a valid input, as has been shown in the case of Adversarial examples \cite{goodfellow2014explaining}. This is why, similarly to the mapping between MNIST and SVHN digits results presented in \cite{taigman2016unsupervised}, a GAN loss is needed. The one-digit-left-out experiments are also very much in line with those presented for the digit mapping when one of the digits is removed, and are, therefore, more indicative of the generalization power of deep networks than on the specific physics problem. An alternative way to GANs to improve generalization may be to rely on activations from multiple layers of the direct network, as is done in the perceptual loss \cite{johnson2016perceptual}. 

Compared to \cite{liu2018generative}, in this work, we utilize our spectra2pix model and showcase the ability of our model to converge without GAN training, and more importantly, we demonstrate a successful generalization ability of the model to design a complete unseen sub-family set of geometries, taken from fairly different distribution the model was trained on. This generalization capability is associated with category two described above.

In \cite{ma2018deep}, the authors introduce a model incorporating two bi-directional networks along with a synthetic dataset composed of vectorized representation of geometries associated with materials, reflection spectra and circular dichroism spectra. The dual bi-directional model comprises two networks, primary network and auxiliary network. The primary network predicts back and forth the geometry encoding vector and its associated reflection spectra Fig. 4 The auxiliary network, predicts back and forth the geometry encoding vector and its associated circular dichroism (CD) spectra. Both networks are separately trained using the dataset above. The authors show that a model that combines both the auxiliary network and the primary network yields more accurate predictions.

In \cite{sajedian2019finding}, the authors suggest a Neural Network that solves the direct problem of inferring spectra for a given geometry. This problem can be solved via (slow) simulations, and is considered to be more feasible compared to the ill-posed inverse problem, of inferring a geometry for on-demand spectra.

\section{Proposed Method}

\begin{figure*}[t]
\includegraphics[width=1.0\linewidth]{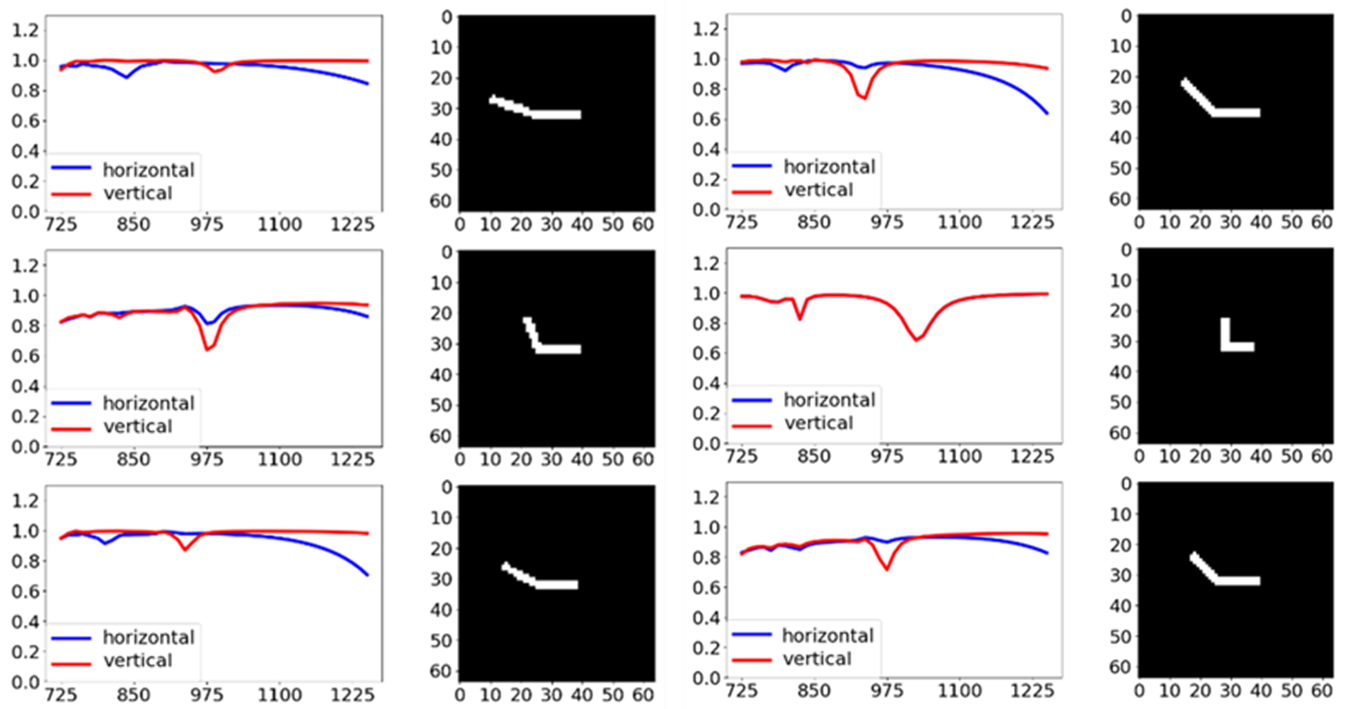}
\centering
\caption{Six samples from the test set showing the variations of the “L” family it includes. None of these variations (shown and unshown) are seen by the spectra2pix during the training phase.}
\label{fig:testset}
\end{figure*}

The section presents the problem setup, the spectra2pix architecture and loss function.

\subsection{Problem Setup}
Let $\mathcal{S}=\left\{s_{i}\right\}_{i=1}^{s}$ be the set of all supported spectra. Let $\mathcal{G}=\left\{g_i\right\}_{i=1}^{g}$ be the set of binary 2D square images of all geometries, where $g_i \in [\left\{0,1\right\}]^{d \times d}$, and $d \in \mathbb{N}$ is the dimension of the images. Each geometry image $g_i \in \mathcal{G}$ is associated with a valid pair of spectra $(s_i^1,s_i^2)$, where $s_i^1,s_i^2 \in S$. Each element in the pairs of spectra is associated with different polarization (vertical or horizontal). Let $C$ be the set of all supported materials. In this study, without loss of generality, we use one material (gold), and a real valued epsilon host $h \in \mathbb{R}$.

We will define $M:S \times S  \times C  \rightarrow G$ to be a model that maps pairs of spectra associated with material properties, into a 2D image that comprises the matched geometry. Given a set of $N$ quadruplet training elements 
\begin{equation}
X = [(s_i^1,s_i^2,c_i,g_i)]_{i=1}^{N}
\end{equation}
our goal is to train a model M such that for all $(s^1,s^2,c,g) \in X$, the generated image 
\begin{equation}
\hat g := M(s^1,s^2,c,g) 
\end{equation}
approximates the label image g with a high accuracy. To this end, we utilize a training procedure that minimizes a loss function, applied between the generated images and the ground truth images.

\subsection{The Loss Function}

Our loss function $L:\mathbb{R}^{d \times d} \times \mathbb{R}^{d \times d} \rightarrow R$ defined as: 
\begin{equation}
L(\hat g, g) = L_{L2} (\hat g, g) = \left\|M(s^1,s^2,c,g)  - g \right\|_{2}^{2}  
\end{equation}
which solely rely on the pixelwise comparison between the generated image and the ground truth image.

By employing a pixelwise loss function on the generated images, our spectra2pix model learns to approximate the hidden inverse function between (1) spectra and material properties and (2) geometries.

\subsection{Model Architecture}

The architecture of spectra2pix is composed of two parts. The first part receives the vectorized representation of the pair of spectra and the material properties as input and apply a set of parallel sequences of fully connected layers. Each sequence of fully connected layers receives a different part of the input data (different polarizations and host material), and utilize a different set of learnt weights. 

The second part of the model architecture, receives the three outputs of the last fully connected layers from the first part, concatenate these three intermediate vectors into one unified representation. The unified vector is then transformed into a higher dimension, by utilizing a fully connected layer. Next, the higher dimensional vector is reshaped into a matrix, and forward through a sequence of three convolutional layers, each followed by a bias and ReLU activations. Each convolutional layer incorporates ten filters with kernel size of $5 \times 5$, except of the last layer which utilizes a single filter. The output of the last convolutional layer is the generated image, denoted by $\hat g$. Then the loss function $L_{L2}(\hat g, g)$  between the ground truth g and the generated image $\hat g$ is applied. The Spectra2Pix model is illustrated in Fig~\ref{fig:architecture}.

\begin{figure*}[t]
\includegraphics[width=1.0\linewidth]{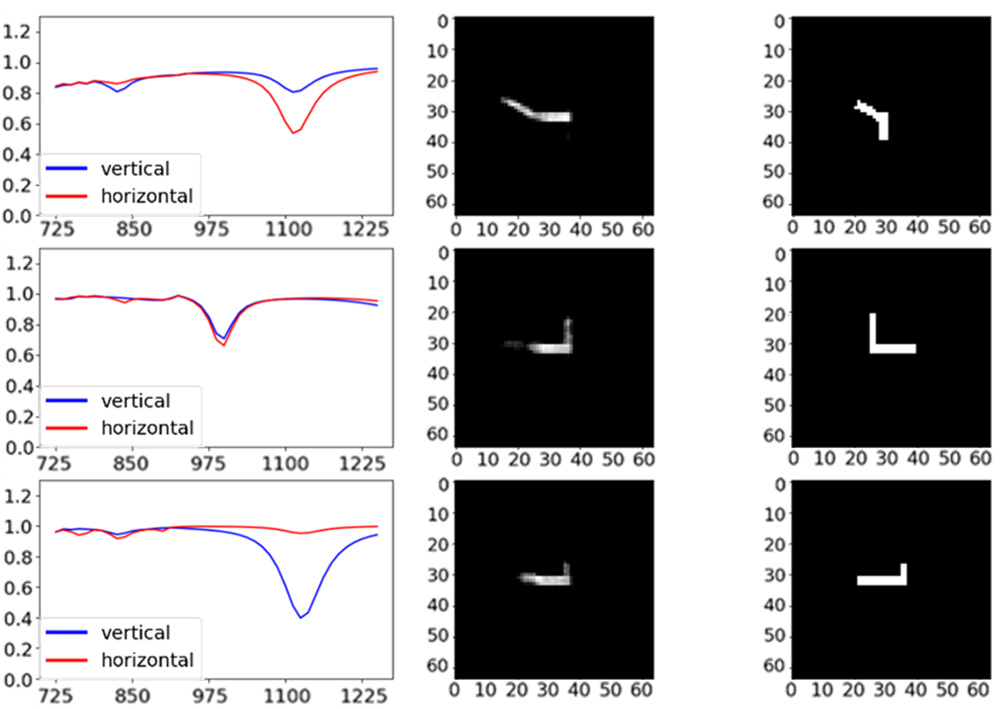}
\centering
\caption{results of queries with three designs from the test set to spectra2pix after the learning phase. On the left column the input spectra are presented, on the middle column we show the predicted geometry by spectra2pix and on the right column the ground truth geometry is depicted. See the main text for discussion on these results.}
\label{fig:test_results}
\end{figure*}

\begin{figure*}[t]
\includegraphics[width=1.0\linewidth]{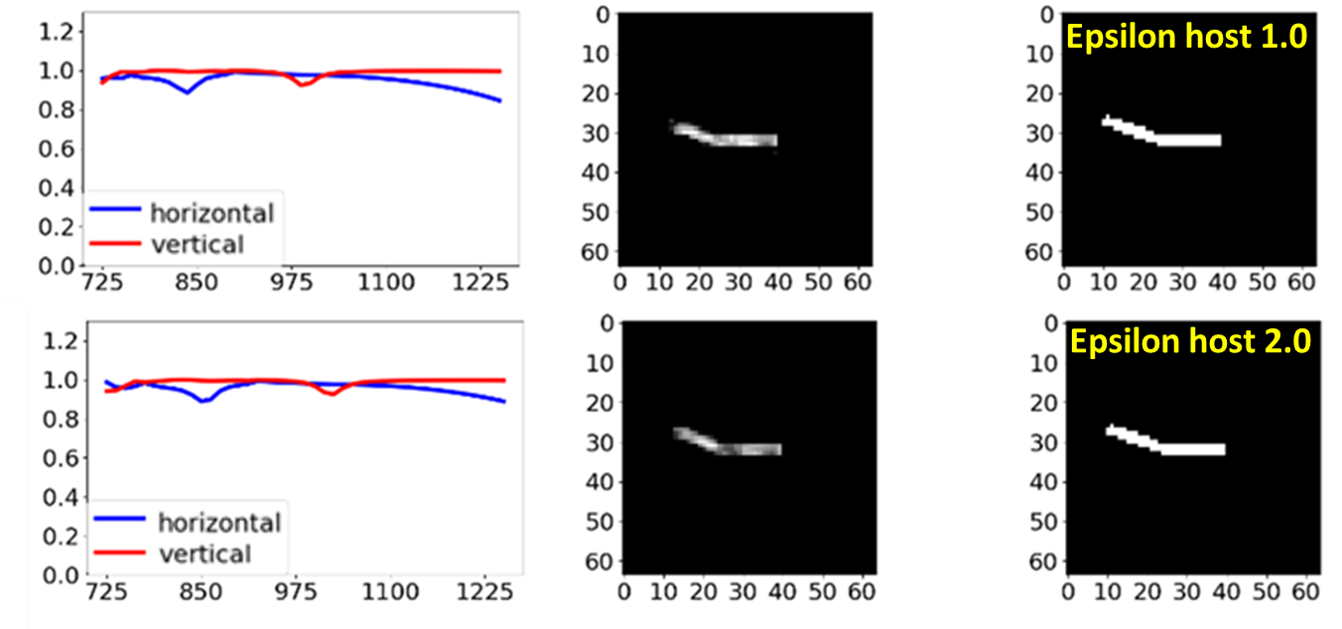}
\centering
\caption{designing the same geometry, with different epsilon host. from left to right: input spectra. generated image, ground truth image. The first and the second row correspond to epsilon host of 1.0 and 2.0, respectively. The model shows consistency in designing the same geometry across different host materials.}
\label{fig:epsilon_host}
\end{figure*}

\section{Experimental Results}

\subsection{The Dataset}
In this work, we utilize the dataset from \cite{malkiel2018deep, malkiel2018plasmonic} . This dataset comprises ~13K samples of synthetic experiments. Each sample associated with a geometry, a single polarization (vertical or horizontal) and material properties. By pairing the polarization, we formed ~6K experiments comprising the quadruplet $(s_1,s_2,c,g) \in X$. The original dataset contains four materials: Gold, Silver, Palladium and Aluminium. Since most of the experiments are associated with Gold or Silver, and since both materials show a strong correlation in their spectra, in this study, we utilize only the experiments where the nanostructures are made of gold, without loss of generality. Nonetheless, we keep the variable values for the host material, each experiment is associated with. The epsilon host dielectric values vary in the range [1.0,3.0]. 

The geometries are composed of different combinations of edges, which together forms a template of the shape “H”. All three data parts, geometry, spectrum and material properties, are represented as vectors. Specifically, for the geometries, an eight-dimensional encoding is used. Five dimensions encode the presence of each one of the five edges of the H shape (binary values). Two dimensions encode the size of (1) the outer edges (which share the same size) and (2) the inner edge. The last dimension represents the angle between the top left outer edge with the inner edge (angles are between 0 to 90).

We transformed the above geometries representation into 2D binary images. A sample of the transformed images, along with the matched spectra of each geometry can be seen in Fig.\ref{fig:dataset}. The transformed dataset is attached as supplementary and is available for the public at  \url{https://github.com/ItzikMalkiel/spectra2pix}.

%{\color{blue} make it anonymous for ICCP}

\subsection{Towards Generalization}
To study the ability of Spectra2Pix to generalize, we split the above dataset into train, test and validation sets. The test set contains all the geometries of the shape L and their variants, i.e. the test contains all L shape geometries with different angles for the top left outer edge, including geometries that are relatively similar to L, such as 'U' with a top left angle that is bigger than 70 degrees. The train set contains all the rest of the experiments, leaving 5\% as a holdout validation set. In summary, the size of the train, validation and test sets used in this study are ~3.3K, ~150, ~700, respectively. A representative sample of test set can be seen in Fig.\ref{fig:testset}.

We train the Spectra2Pix network for 1M training steps, with a batch size of 64. Adam optimizer is being used with a learning rate of $1e^{-5}$, $\beta_{1} = 0.9$, $\beta_{2} = 0.999$, and $\epsilon = 1e^{-8}$. We use the validation set for early stopping. 

At the end of the training, we used the model to infer geometries for the test set. Figure \ref{fig:test_results} exhibits a representative sample from the test set predictions. Each row represents a different query. The left column exhibits the input spectra (both vertical and horizontal polarization), the middle presents the generated geometry $\hat g$ and the right showcase the ground truth label g. For the first row, a spectra of L shape geometry, along with an epsilon host of 1.0, was fed into the Spectra2Pix model. The model generated an image of L shape with a somehow similar size and angle, but with different orientation. In the second row, the model was able to generate an L geometry with a similar size and angle, but with a symmetric pose, which does not affect the spectra. In addition, for this sample, and since we plot the raw values of the generated image, it can be seen that the model is not confident enough about the size of the bottom edge, as some artifacts are presented in the left side of the generated image. For the third experiment, the model was able to infer a fairly accurate L shape. Overall, these results showcase the ability of spectra2pix to generalize to unseen geometries sampled from a fairly different distribution the model was trained on.

\subsection{Learning host material}
Figure \ref{fig:epsilon_host} showcase the ability of Spectra2Pix model to learn the dependence of the epsilon host material. In this figure, we queried the network with two different pairs of spectra that are associated with the same geometry but with different epsilon host material. The first pair corresponds to a “L” geometry, hosted in an epsilon dialectic of 1.0. The second pair associated with an identical geometry, but hosted in an epsilon dielectric of 2.0.

\subsection{Limitations}
\begin{figure*}[t]
\includegraphics[width=0.8\linewidth]{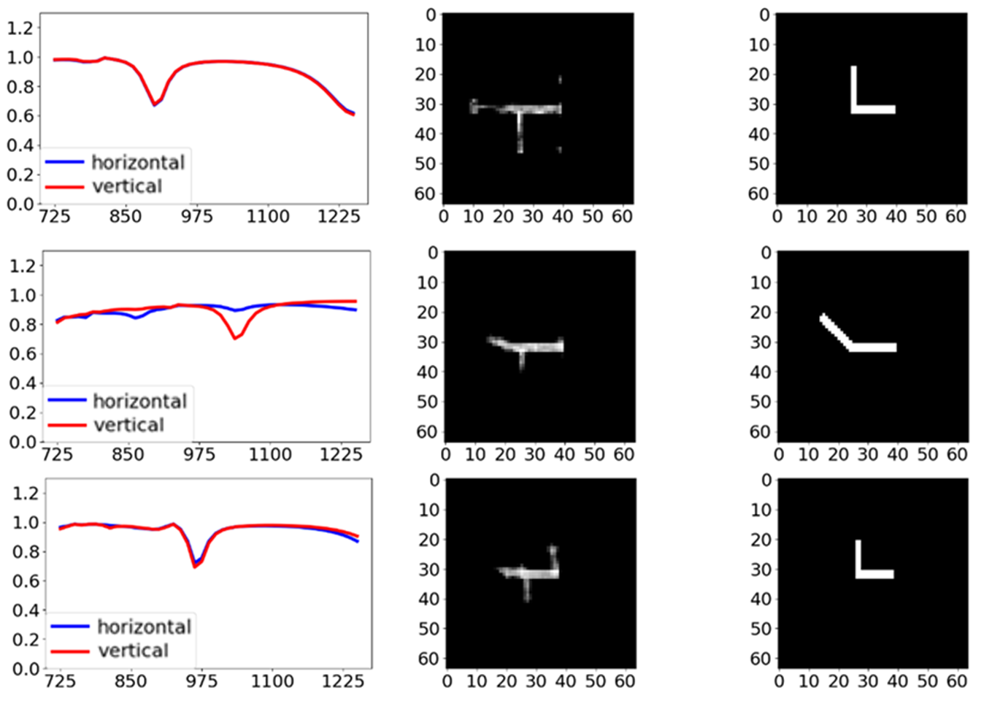}
\centering
\caption{Three representative failure cases as described in the main text.}
\label{fig:limitations}
\end{figure*}

Additionally, we explored some of the limitation of the above method and dataset. Figure \ref{fig:limitations} showcase three representative samples of geometries for which the model produced non-optimal designs.

The observed discrepancies can be categorized into three classes. The first corresponds to low confidence of the model, for the existence of some edges. For example, the first row in Fig. \ref{fig:limitations} presents a design of L shape, where the model was able to infer somehow a relatively solid design for a symmetric L (which shares the same spectra as the ground truth), but the generated image comprises artifacts in few places, especially in the area of the continuation of the inner edge upon the left direction, which looks like an extra edge which form a superposition of two L shapes (each one is a mirror of the second one). The second class relates to the existence of extra small edges. For example, in the second row of Fig.\ref{fig:limitations}, the model was able to generate a similar L shape, somehow with a non-accurate angle and an extra small edge at the bottom. The extra small edge at the bottom has a negligible affect on the spectra for both polarizations, and might compromise for the discrepancy in the angel of the top left outer edge of the predicted geometry. %This specific example also exhibits the ability of the model to yield outer edges of different sizes, that cannot be achieved in the architecture of the bi-directional netwrok from \cite{malkiel2018deep, malkiel2018plasmonic} .
The third class incorporate failure cases where the model designs a different geometry than intended. These designs should follow a verification using simulations, since they might yield similar spectra to the input spectra, although they share different shape compared to the ground truth geometry. Alternatively, a bi-directional model can shade light on the verification of such designs. An example of this category can be seen in the third row of Fig.\ref{fig:limitations}, for which, it also seems that the model predicted a superposition of two 'L' shapes.

%failure cases. The first row showcase a generated image that suffers from artifacts, which correlates with low confidence for the existence of some edges. Second row exhibit a deisgn comprising of and extra small edge, which have a negligible affect the spectra. The third row presents a wrong prediction that looks like superpi, which requires a verification by simulations, since it might share a similar spectrum. {\color{blue} fix me}

We attribute the above discrepancies to: (1) the complexity of the underlying physical process.  (2) The difficulty to generalize to completely unseen geometries, given a relatively smaller-sized training set comprising only eight different shapes with 4K variants (where a big portion of the samples give emphasize to the different variants of epsilon host, rather than the richness of the geometries). (3) The existence of multiple valid $M:S \times S \times C \rightarrow G$ functions, since the same spectra and material properties can be matched with multiple geometries. For example, for some geometries such as L shapes, flipping the geometry doesn’t affect the spectra. 

The second difficulty mentioned above can be solved by (A) the utilization of a larger and richer dataset, that would ease the generalization and robustness of such a design model, or by (B) leveraging a bi-directional architecture, which utilizes an extra model (say “pix2spectra” architecture) that predicts back the matched spectra of the generated image. The bi-directional model can regularize the training, and encourage the spectra2pix model to produce images with higher confidence and less artifacts, since given an accurate pix2Spectra model, image artifacts and low confidence of the existence of edges would yield higher penalty in the predicted spectra. 

The third difficulty above, which indicates that the hidden function between spectra and geometry is not a well-defined function, can be solved by leveraging Generative Adversarial Networks (GANs). A GAN based model, incorporating a discriminator network, may be able to detect low confidence of existence of edges, image artifacts and superposition of the same edge, as generated image, which will then encourage the generator model to avoid such behavior. GANs can also be used to produce a set of different geometries that matches a single input spectra. In this work we leave the above for farther investigation.

%\subsection{Discussion}

%{\color{blue} report quantitative results. run leave one out. emphasize that in (our) previous work didn't do this generalization test. Compare results with previous work }

\section{Conclusion}
The use of machine learning techniques and deep learning in particular has spawned huge interest over the past few years in the nanophotonics communities due to the great promises these techniques offer for the inverse design of novel devices and functionalities. In this paper, we introduce spectra2pix, a model comprising of ultimate degrees of freedom, that conceptually allows the design of any 2D geometry. In addition, we present the ability of spectra2pix to successfully generalize for the designing a set of completely unseen sub-family of geometries. Our results highlight the importance and the generalization ability of Deep Neural Networks, towards the goal of inverse design of any nanostructure with at-will spectral response. To our knowledge, and compared to other work in the field, spectra2pix is the first model to present a generalization ability of designing a completely unseen sub-family of geometries sampled from a fairly different distribution the model was trained on. 
% WE would like to encourage the comuunity to conduct similar evaluations and adopt our notaition for generalization levels, when measuring the performance of various design model.

% Any acknowledgments to only be included in camera ready
% \ifpeerreview \else
% \section*{Acknowledgments}
% The authors would like to thank...
% \fi
% \clearpage
\bibliographystyle{IEEEtran}
\bibliography{references}

\ifpeerreview \else
%%%% For the camera ready version, please fill out this
%%%% biography. Your camera ready should be within a 12 page limit
%%%% including acknowledgments, references and biography.

% If you have an EPS/PDF photo (graphicx package needed) extra braces are
% needed around the contents of the optional argument to biography to prevent
% the LaTeX parser from getting confused when it sees the complicated
% \includegraphics command within an optional argument. (You could
% create your own custom macro containing the \includegraphics command
% to make things simpler here.)
% \begin{IEEEbiography}[{\includegraphics[width=1in,height=1.25in,clip,keepaspectratio]{mshell}}]{Michael Shell}
% or if you just want to reserve a space for a photo:

% \begin{IEEEbiography}{Michael Shell}
% Biography text here.
% \end{IEEEbiography}

% insert where needed to balance the two columns on the last page with
% biographies
%\newpage

% if you will not have a photo at all:
% \begin{IEEEbiographynophoto}{John Doe}
% Biography text here.
% \end{IEEEbiographynophoto}

% You can push biographies down or up by placing
% a \vfill before or after them. The appropriate
% use of \vfill depends on what kind of text is
% on the last page and whether or not the columns
% are being equalized.
%\vfill

\fi

\end{document}